# Comments on "Improving Transferability Between Different Engineering Stages in the Development of Automated Material Flow Modules"

Kleanthis Thramboulidis
Electrical and Computer Engineering, University of Patras, Greece.

*Abstract*—In the paper by D. Regulin *et al.* (*IEEE Trans. On Automation Science and Engineering, vol. 13, no. 4, 1422-1432, October 2016*) authors claim that they present a meta-model for the modeling of the Automated Material Flow Module (aMFM) and a model-driven design approach for aMFMs. In this letter we comment on the presented meta-model and the proposed model-driven approach regarding their potential for exploitation. We present specific arguments and make cases that call the authors design decision into question.

*Index Terms*—Industrial automation systems, model-driven development, meta-modeling.

## I. INTRODUCTION

For the definition of an efficient model-driven development process the source and the target models as well as the automation of the transformation process of the source model to the target one are required. Meta-models are usually used to describe the source and the target models. Authors claim in [1] that they present: a) a meta-model, namely AutoMFM, for the modeling of an automated Material Flow Module (aMFM) and b) a process to facilitate the development of Automated Material Flow Systems (aMFSs) which are considered as compositions of aMFMs.

In this letter, the proposed in [1] meta-model is discussed in the context of the model-driven development paradigm. Specific arguments that call into question the authors' decisions and claims, regarding the maturity, robustness and effectiveness of the AutoMFM, are presented. Taking into account that a rigorous process assumes a robust, mature and effective meta-model, we present only a few comments on the process presented in [1]. The remainder of this letter is organized as follows. Section II establishes the context of the discussion which is presented in Section III.

## II. THE CONTEXT OF THE DISCUSSION

Model-driven engineering (MDE) has been successfully used in the manufacturing domain to alleviate the complexity of platforms and express domain concepts effectively [2]. Models are used to represent in a formalized way the structure, the behavior, and the requirements of the system under development. Several approaches focused not only on the software discipline of manufacturing system but also on the system level where the system is considered as a composition of mechatronics/cyber-physical components, e.g., [3].

The objective of constructing a model for a system, and thus for an aMFM, is to address its complexity by describing it in an abstract way. This model should be independent of the final execution platform and thus it has to be transformed to an executable one. The automation of this transformation leads to a model-driven development process. The initial, platform independent model is known as the source model and the final executable one is known as the target model. A model for an aMFM would be a description of the aMFM at the system level expressed in a well-defined language. The meta-model is used to model/describe this well-defined language. Thus, system models constructed using a specific meta-model conform to the meta-model and this check is automated in a model-driven development process. Moreover, meta-models are used to formalize the domain knowledge and facilitate the job of the system engineer [2]. Obviously this should be the intent of the AutoMFM but this is not as it is argued in this letter.

For the development of a multi discipline system, such as manufacturing systems, it is common to use several tools and models for the various disciplines. AutomationML, as is also the case for ISO10303-233 [4], defines a standardized neutral data exchange format based on XML for the storage and exchange of plant engineering information among the various discipline tools involved in the engineering of the plant.

## III. DISCUSSION

Authors claim in [1] that the proposed meta-model and the approach have specifically developed for the domain of aMFSs. However, it is not clear what makes the proposed meta-model specific for aMFSs since the only domain specific information captured in the meta-model is the types of logistic tasks [1, Ref. 31]. This is also evident from [1, Sec. II] and [1, Ref. 31] where none of the defined requirements is specific for aMFSs. Arguments should be given to claim that material flow modules cannot be effectively modeled by an industrial automation domain modeling language. Moreover, the benefits of such a very specific modeling language have to overwhelm the cost of developing and maintaining another language. The case of extending an industrial automation domain language to address the specific requirements of material flow systems (if these exist) is the obvious way to go.

Authors claim that the paper presents a methodology for a data exchange between the different tools applied during the engineering cycle in the design process of aPS [1, Note to Practitioners]. However, it is evident that the objective of the AutoMFM is to provide the architecture of the system. This is claimed by authors in several places, e.g., [1, Sec. I] where authors claim that the system architecture is provided by the metamodel, and [1, Ref.31] where authors emphasize that the presented meta-model is their solution for a model-based module description [1, Ref.31] in order to "handle the complexity and, thereby, reduce the effort regarding development in functional engineering." It is also clear that [1] does not describe a methodology for data exchange between tools. Thus, the AutoMFM is considered in this letter as a model of the modeling language that is used to model the aMFM to facilitate the development process of aMFSs. AutoMFM fails to address this requirement. A comparison of the AutoMFM as presented in Fig.1 [1, Ref.31, Fig.1] and the model of the Crane module, which is based on the AutoMFM and is given in Fig.2 [1, Ref.31, Fig.6] is an initial indication of this claim. AutoMFM does not define a modeling language neither a robust system partitioning as it is argued in the following.

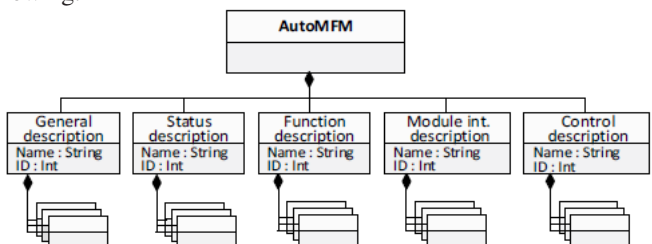

Fig. 1. The AutoMFM [1, Ref.31, Fig.1].

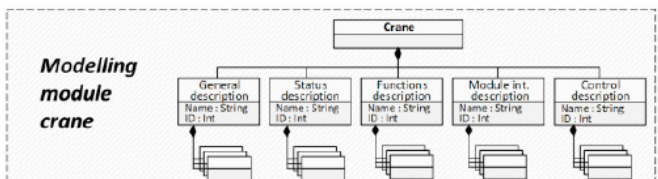

Fig. 2. The Model of the crane module [1, Ref.31, Fig.6].

Authors do not refer the notation that is used to represent the model. The only reference to the modeling notation is given in the application example [1, Sec.V], where authors state that "The logistic functionality and behavior description of the T-junction was described in a Pert chart diagram."

Since this paper is based on AutoMFM which is briefly described in [1] the rest of the letter is based also on the detailed description of the meta model that is given in [1, Ref. 31].

*A. Partitioning as a mechanism to address complexity*

Authors claim in [1] that the setup of a modular software architecture, where each aMFM can execute specific tasks and interact with neighboring modules is their solution, i.e., contribution, to minimize the static structure of the aMFS and reduce the effort and error-proness [1, Ref. 31, Sec. I]. Moreover, they claim that the motivation for that is the fact that reuse of control software is currently applied through copy paste [1, Ref. 31, Sec. I]. Partitioning as a mean for handling complexity is already known for many years and is exploited by almost any development approach which model the system as a composition of parts. Reusability is also exploited to a great extend by various mechanisms even in the domain of industrial automation software, e.g., function block libraries.

Authors define the aMFM as the main unit of composition of an aMFS and provide a meta model for this, i.e., the AutoMFM. A complete modeling notation for modeling an aMFS should define not only the model of the unit of composition, but also the means of integrating these units as well as the means of modeling the interaction of the aMFS with its environment. Based on that, AutoMFM cannot be considered as an autonomous modeling language for aMFSs. Moreover, its effectiveness for modeling the aMFM is questionable.

One may partition a complex system in various ways. For example, the procedural paradigm defines the function as the unit of partitioning, while the object-oriented defines the object. Authors define, through the AutoMFM, that the model of the aMFM is constituted from five parts that correspond to the (sub)classes of the AutoMFM, as shown in Fig. 1. Based on these parts, authors define the architecture of the module. However, authors do not argue on the reasoning of the proposed partitioning in terms of software and system engineering, not even present the benefits of such a partitioning. Moreover the proposed partitioning raises many questions that are discussed in the following.

Authors propose the separation of constant properties of a module from its variable ones. The only argument for this is that constants are particularly essential during engineering and planning. However, the functionality of the module and its state, which is expressed through variables, are also essential during these activities of the development process. Authors also include in this part which is defined by the General (sub)class, information on the current location of movable components [1, Ref. 31, Sec 4.2], which is variable. Moreover, this decision destroys the cohesion of the part.

Surprisingly, information on orders of customers, current and future ones, is included in the part of the module that captures the status related information of the module [1, Ref. 31, Sec 4.3].

The partitioning of the Control description module of the aMFM into control function and global variable list [1, Ref. 31, Fig. 3] is inconsistent with the claim that the proposed aMFM model captures the architecture of the aMFM.

Authors define a part of the model of the aMFM to capture the function description, while at the same time they have another part, i.e., the control description to capture the control functionality of the module, which captures part of the module's behavior [1, Ref. 31, Sec 4.5]. It is not clear if the function part of the model contains specification of the supported functionality types or specification of the behavior of the aMFM. It should be noted that at this level of partitioning, the behavior of the component should be specified independent of the implementation technologies, i.e., mechanics, electronics and software. The decision of splitting the behavior follows and such a partitioning should be effectively supported by the meta model and this is not the case with AutoMFM.

The decision of authors to include in the model of the aMFM information regarding material flow between different modules [1, Ref. 31, Sec 4.5] is highly restrictive to the reusability of the model.

The importance of interfaces is well known in the software and system engineering domain for many years now. There is no need to have a survey to argue that module interfaces are a critical issue in the realization of a system [1, Ref.31, Sec.2]. To satisfy this requirement, authors propose to model the interfaces of the module in the Module interfaces description (sub)class [1, Ref. 31, Sec 4.6] without giving any mechanism or formally defining the association of this part with the other two parts that capture behavior, i.e., Function description and Control Description. Input and output information of a large number of control functions which are modeled in the Control description class constitute part of the module interface, i.e., module Interface (sub)class and this introduces a very high coupling between the corresponding classes. Moreover, authors claim that modeling the classes module interface description and control description depend on the neighboring modules [1, Ref.31, Sec.5.2]. This leads the authors to include in the models of both modules this information, introducing redundancy in the model as well as a high coupling between the modules. Authors claim that in this way both modules contain the same consistent interface. The proposed solution, at least in the way it is described, is far behind state-of-the-art in Software and System Engineering. Software Engineering has successfully addressed effectively this requirement e.g., provided and required interfaces.

By applying the proposed partitioning authors capture: a) the constant properties of the Conveyor aMFM [1, Ref.31, Sec. 5] by the General description (sub)class, b) its properties that change during run-time by the Status description (sub)class, c) its functionality by the Function (sub)class), d) the description of the software and the hardware that controls the Conveyor by the Control description class and finally model interactions of the Conveyor with neighboring aMFMs with the Interface (sub)class. Since a different definition of the term class is not given in [1] it is assumed that the well known meaning of class is adopted. Given that, the proposed decomposition of the structural and behavioral characteristics of a class (the one that represents the aMFM) in the above classes needs at least an analysis regarding various aspects involved in system and software modeling, otherwise it is completely arbitrary.

### B. The AutoMFM

Even though authors claim that the aMFM meta model enables the composition of multiple modules to a system module [1, Ref. 31, Sec 1] this is not described. Moreover, the modular approach in system development has been applied in industrial automation for years now and based on this authors should argue on the benefits that their meta-model introduces in the system composition.

Authors claim that in order to improve clarity and applicability of the AutoMFM, they do not directly allocate attributes and functions to the class to model information and behaviour respectively [1, Ref. 31, Sec. 4.1]. Firstly, this is not true since almost all (sub)classes in the meta model appear with a few attributes and secondly, it is not possible to define a class without defining structural and behavioral properties of its instances. Furthermore, even though authors argue that attributes model information and functions model behavior they claim that the presented partitioning into (sub)classes is done to separate module information into (sub)classes [1, Ref. 31, Sec. 4.1]. What about the behavior?

Traceability is not facilitated, as authors claim [1, Ref. 31, Sec. 4.1], by assigning a name and a globally unique identifier to a class.

Authors claim that they model in AutoMFM the control description to fulfill two requirements: a) generation of software instances (R4) and b) support of different operating strategies (R1) [1, Ref. 31, Sec. 4.4]. The first requirement is supported by any development approach. The second is supported by the most of the approaches already used in practice. Thus, the contribution of the proposed approach regarding these two requirements is not given.

The claim of the authors that logical function of the aMFM written in the programming languages of the IEC61131-3, can be modelled in the (sub)class Control description of the AutoMFM [1, Ref. 31, Sec. 4.4] is completely inconsistent with the definition of the AutoMFM as a meta-model, i.e., a model that defines the modeling language for modeling the aMFM.

### C. Development Process

Authors use the term stage inconsistently. In some places they use the term to refer to different engineering activities such as module planning or functional [1, Introduction]. In other places to refer to specific time periods inside a stage, e.g., "different design stages, i.e., time periods" [1, Note to practitioners] and "the whole development process of the aPS is divided into stages, i.e., time periods" [1, Sec. I]. Moreover, plant planning, functional engineering, and commissioning/production are referred as planning stages, e.g., "planning stages, i.e., plant planning, functional engineering, and commissioning/production" [1, Sec. II], which is inconsistent with Fig. 2 [1].